\documentclass[a4paper,11pt]{article}
\pdfoutput=1 

\usepackage{jheppub} 

\newcommand{\Equation}[2]{  \begin{equation}\label{#1}#2\end{equation} }

\newcommand{\SubAlign}[2]
{\begin{subequations}\label{#1}\begin{align}#2\end{align}\end{subequations}}

\newcommand{\mbf}{\mathbf}
 
\newcommand{\bs}{\boldsymbol}
\newcommand{\Figref}[1]{Fig.~\ref{#1}}
\newcommand{\Eqref}[1]{\eqref{#1}}
\newcommand{\Real}{\mathbb{R}}

\newcommand{\Grad}{{\bs \nabla}}
\newcommand{\Div} {{\bs \nabla}\!\cdot\!}

\newcommand{\ScalarProd}[2]{\left\langle #1,#2\right\rangle}
\newcommand{\x}{\mathbf{x}}

\newcommand{\ez}{{\bs e}_z}

\newcommand{\A}{{\mbf A}}

\newcommand{\Op}{\mathcal{O}}
\newcommand{\U}{\mathcal{U}}
\newcommand{\vL}{\mathbf{L}}

\newcommand{\la}[1]{\mbox{$
\lefteqn{ \mbox{\,\, \tiny #1}}$} \label{#1}}
\renewcommand{\la}[1]{\label{#1}}
\usepackage[usenames,dvipsnames]{color}	
\usepackage{xcolor}			
\newcommand{\Red}    [1]{{\color{red}{#1}}}

\newcommand{\Purple} [1]{{\color[rgb]{0.5,0.0,1.0}{#1}}}
\newcommand{\Olive}  [1]{{\color[rgb]{0.5,0.5,0.0}{#1}}}

\title{\boldmath Vortex precession and exchange in a  Bose-Einstein condensate}

\author[a]{Julien Garaud\,}
\author[b]{Jin Dai\,}
\author[b,c,d]{Antti J. Niemi}

\affiliation[a]{Institut Denis Poisson CNRS-UMR 7013, Universit\'e de Tours, 37200 France}
\affiliation[b]{Nordita, Stockholm University and Uppsala University, \\  
				Roslagstullsbacken 23, SE-106 91 Stockholm, Sweden}
\affiliation[c]{Pacific Quantum Center, Far Eastern Federal University, \\
				690950 Sukhanova 8, Vladivostok, Russia}
\affiliation[d]{Department of Physics, Beijing Institute of Technology, 
				Haidian District, Beijing 100081, China}
\emailAdd{garaud.phys@gmail.com}
\emailAdd{jin.dai@fysik.su.se}
\emailAdd{Antti.Niemi@su.se}

\abstract{
Vortices in a Bose-Einstein condensate are modelled as spontaneously symmetry breaking  
minimum energy solutions of the time dependent Gross-Pitaevskii equation, using the 
method of constrained optimization. 
In a non-rotating axially symmetric trap, the core of a single vortex precesses around 
the trap center and, at the same time, the phase of its wave function shifts at a 
constant rate. 
The precession velocity, the speed of phase shift, and the distance between the vortex 
core and the trap center, depend continuously on the value of the conserved angular 
momentum that is carried by the entire condensate. 
In the case of a symmetric pair of identical vortices, the precession engages an emergent 
gauge field in their relative coordinate, with a flux that is equal to the ratio between 
the precession and shift velocities.
}


\begin{document} 
\maketitle
\flushbottom

\section{Introduction}
\label{Sec:intro}
 

Bose-Einstein condensate is a widely investigated realization of a coherent macroscopic 
quantum state. In particular, dilute condensates of trapped ultra-cold atoms have unique 
quantum features that facilitate a high level of experimental control \cite{Anderson-1995,
Davis-1995,Bradley-1997}. Various realizations are studied vigorously, both in earth-bound 
and in earth-orbiting laboratories \cite{Aveline-2020}. Among the goals is the development 
of ultra-sensitive sensors and detectors \cite{Elliott-2018}, and the properties of cold 
atom condensates are also employed as a platform for quantum computation and simulation 
\cite{Bloch-2012}.

%
%
%
%
%
%
%
%
%
%
\begin{figure}[ht]
\centering 
\includegraphics[width=.75\textwidth,origin=c,angle=0]{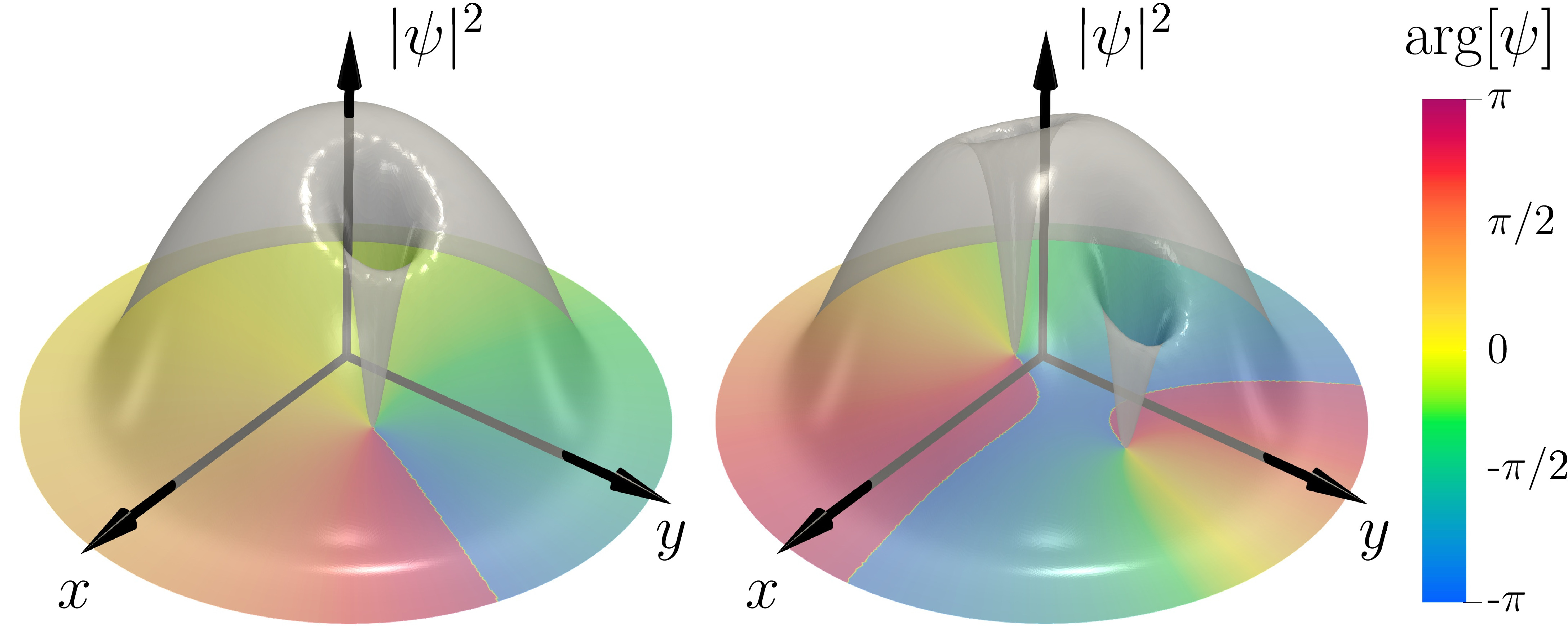}
\caption{ 
Examples of asymmetric vortex configurations of a Bose-Einstein condensate in an 
axially symmetric harmonic trap.
The elevation of the semitransparent surface stems for the condensate density $|\psi|^2$, 
while the coloring projected onto the $xy$-plane indicates the value of the phase 
$\arg[\psi]$ of the macroscopic wave function. 
The examples shown are minima of the two dimensional Gross-Pitaevskii free energy 
\Eqref{H}, with non-integer values of the angular momentum \Eqref{L} $L_z=0.80$ 
(left panel) and $L_z=1.48$ (right panel). 
The vortices precess around the trap center in accordance with the time dependent 
Gross-Pitaevskii equation \Eqref{GPeq}. 
}
\label{Fig:Vortex:0}
\end{figure}
%
%
%
%
%
%
%

Quantum vortices are the principal topological excitations in a cold atom Bose-Einstein 
condensate \cite{Matthews-1999,Chevy-2000,Raman-2001,Abo-Shaeer-2001}. A vortex is 
characterized by an integer valued circulation of the macroscopic wave function 
$\psi(\x)$
\begin{equation*}
\frac{1}{2\pi}\oint d{\bs\ell}\!\cdot\!\Grad\arg[\psi]  \in \mathbb Z
\end{equation*}
around its core. The time-dependent Gross-Pitaevskii equation \cite{Gross-1961,
Pitaevskii-1961} governs the dynamics of the macroscopic wave function, at the level 
of the mean field theory. This is a nonlinear Schr\"odinger equation with a quartic 
nonlinearity that accounts for interactions between the atoms \cite{Pitaevskii-2003,
Lieb-2006,Pethick-2008,Fetter-2009,Bao-2013}. 

Vortices that appear in rotating cold atom Bose-Einstein condensates have been extensively 
studied, both experimentally and theoretically \cite{Haljan-2001,Fetter-2001,Fetter-2009}. 
At the level of the Gross-Pitaevskii equation these vortices are commonly modelled 
as stationary solutions in a co-rotating frame \cite{Butts-1999,Aftalion-2001,
Seiringer-2002,Pitaevskii-2003,Lieb-2006,Pethick-2008,Fetter-2009,Bao-2013}.
The rotating condensate accommodates growing angular velocity by forming vortices. 
This goes with a discontinuous increase in angular momentum: A rotating 
condensate does not support an arbitrary value of the angular momentum \cite{Butts-1999}.  

Here we are interested in solutions of the Gross-Pitaevskii equation in a trapped 
condensate that does {\it not} rotate. In that case, the equation has no time independent  
solutions besides the trivial one $\psi(\x) \equiv 0$. However, we show that whenever the 
non-rotating condensate supports a non-vanishing {\it arbitrary} valued angular momentum, 
there is a stable minimum energy solution of the time {\it dependent} Gross-Pitaevskii 
equation. These solutions describe eccentric vortices that precess around the center of 
the non-rotating axially symmetric trap, as illustrated in \Figref{Fig:Vortex:0}. 
In particular, the single vortex solutions are very similar to the experimentally 
observed precessing vortices reported in \cite{Anderson-2000}.

It is notable that even though the solutions that we present are minima of the ensuing 
free energy, they are not critical points of the free energy. Thus we employ methods 
of constrained optimization \cite{Fletcher-1987,Nocedal-1999}, to solve the time dependent 
Gross-Pitaevskii equation.

We note that the {\it microscopic}, individual atom angular momentum is certainly 
quantized. But as pointed out in \cite{Mottelson-1999,Kavoulakis-2000} (see also 
\cite{Butts-1999}), in the limit where the number of atoms is very large, the condensate 
can accommodate arbitrary values of the {\it macroscopic} angular momentum that appears 
as a Noether change in the Gross-Pitaevskii equation. 
It was pointed out in \cite{Mottelson-1999,Kavoulakis-2000} and confirmed by 
the variational analysis in \cite{Kavoulakis-2000}, in the case of a single vortex,
that a solution  with arbitrary angular momentum must in general be time dependent.
Our results confirm this proposal.

We emphasize the following difference between our approach that considers a non-rotating 
trap and builds on the ideas of \cite{Mottelson-1999,Kavoulakis-2000}, and the more 
common approach that considers vortices in a rotating trap, in terms of a co-rotating 
frame \cite{Butts-1999,Aftalion-2001,Seiringer-2002,Pitaevskii-2003,Lieb-2006,
Pethick-2008,Fetter-2009,Bao-2013}: 
In the rotating trap problem, the minimal energy state is controlled by the value of 
the angular velocity of the trap that appears as a free parameter in the stationary, 
time independent co-rotating Gross-Pitaevskii energy functional. The vortices appear as 
critical points of this energy functional, as solutions of the time {\it independent} 
Gross-Pitaevskii equation they minimize an unconstrained variational problem.  
In contrast, in our case we use the fact that the angular momentum is conserved, 
thus we fix its value. The vortices are then solutions to a constrained optimization 
problem and solve the time {\it dependent} Gross-Pitaevskii equation. The difference 
between these two problems is detailed in Appendix~\ref{Sec:Comparison}.

Since the vortex configurations that we find are simultaneously both constrained minima 
of the Gross-Pitaevskii free energy, and time dependent solutions of the ensuing 
Hamilton's equation, they bear a certain resemblance to the concept of a classical 
Hamiltonian time crystal \cite{Shapere-2012,Wilczek-2012}. 
Our approach is modelled on the general theory developed in \cite{Alekseev-2020}; 
see also \cite{Dai-2020}. 

We shall further observe that in the case of a symmetric pair of two identical vortices, 
the evolution of the minimal energy vortex states, at specified angular momentum, 
brings about an exchange of the vortices. The exchange of identical bodies in two 
dimensions have been studied extensively, as it can be used to identify anyons. 
Anyons are two-dimensional quasiparticles that are neither fermions nor bosons 
\cite{Leinaas-1977,Wilczek-1982}. The anyon statistics implies that the exchange of 
the position of two identical particles results in a change of their quantum mechanical 
phase which differs from 0 (the case of bosons) or $\pi$ (the case of fermions); 
for reviews on anyons see {\it e.g.} \cite{Iengo-1992,Lerda-1992}). The anyons, that 
can play an important role for quantum computing \cite{Kitaev-2003}, have recently 
been observed by electronic interferometry in quantum Hall effect experiments 
\cite{Nakamura-2020}.

\section{Theoretical developments}

We consider a two dimensional single-component Bose-Einstein condensate on the $xy$-plane 
with an axially symmetric, non-rotating, harmonic trap. This approximates for example an 
anisotropic three dimensional trap, resulting in an oblate condensate. Following 
\cite{Lieb-2002,Lieb-2006} we assume that the condensate describes a dilute limit of a 
large number of atoms,  with a two-body repulsive interaction potential between the atoms.
The condensate is described by a macroscopic wave function $\psi(\x,t)$ 
whose dynamics obeys the (dimensionless) time dependent Gross-Pitaevskii equation 
\cite{Lieb-2002,Lieb-2006}: 
\begin{equation}
i\partial_t\psi=-\frac{1}{2}\nabla^2\psi +\frac{|\x|^2}{2}\psi+g|\psi|^2\psi 
\equiv \ \frac{\delta F} {\delta \psi^\star} \,.
\la{GPeq}
\end{equation}
Here $g$ is a dimensionless coupling that accounts for the interactions between the atoms. 
In a typical Bose-Einstein condensate there are $N_a\sim 10^4-10^6$ ultracold alkali atoms
and the corresponding  coupling is typically $g\sim1$--$10^4$; for discussion 
of the nondimensionalization, see, e.g, \cite{Bao-2013} and the 
Appendix~\ref{Sec:Dimensionless}. The free energy $F$ in eq.~\Eqref{GPeq} is
\begin{equation}
F=\int d^2x \left\{\frac{1}{2}|\Grad\psi|^2 +\frac{|\x|^2}{2}  |\psi|^2 
+\frac{g}{2}|\psi|^4\right\}\,.
\la{H}
\end{equation}

We observe that $F$ is a strictly convex functional, it can not have any critical points 
besides the global minimum $\psi \equiv 0$. Thus there are no nontrivial time 
{\it independent} solutions of the Gross-Pitaevskii equation \Eqref{GPeq}, only time 
dependent nontrivial solutions of \Eqref{GPeq} can exist.
\footnotemark\footnotetext{
In the case of a rotating condensate \cite{Butts-1999,Aftalion-2001,Seiringer-2002,
Pitaevskii-2003,Lieb-2006,Pethick-2008,Fetter-2009,Bao-2013} the angular velocity makes 
the energy into a non-convex functional with non-trivial critical points that describe 
vortices with a discrete valued angular momentum.}
Provided appropriate conditions can be introduced, these solutions can also minimize the 
free energy $F$ in the framework of constrained optimization. Thus, we search for time 
dependent solutions of \Eqref{GPeq} that are also minima of the free energy \Eqref{H}, 
where the latter is subject to an appropriate condition.

Indeed, besides the free energy \Eqref{H} the time evolution equation \Eqref{GPeq} 
supports two additional conserved quantities, as Noether charges: 
The macroscopic angular momentum along the $z$-axis 
\begin{equation}
L_z  \ \equiv \ \int d^2x\, \psi^\star\, \ez\cdot(- i\hbar\,\x\wedge \Grad)\psi \,
\la{L}
\end{equation}
and the norm of the macroscopic wave function {\it a.k.a.} the number of particles 
\begin{equation}
N \ \equiv \ \int d^2 x \, \psi^\star \psi   \,.
\la{Q}
\end{equation}
Accordingly, we search for solutions of the time dependent Gross-Pitaevskii equation 
\Eqref{GPeq}, with fixed values of both the angular momentum $L_z$ and the number of 
particles; in the following, with no loss of generality, we normalize the wave function 
so that $N=1$ (see Appendix~\ref{Sec:Dimensionless} for details). 
On the other hand, the angular momentum  \Eqref{L}, can assume arbitrary values $L_z=l_z$, 
and even in the case of a single vortex the values $l_z$ of the angular momentum are not 
expected to be integer valued \cite{Mottelson-1999,Kavoulakis-2000}.

The Lagrange multiplier theorem \cite{Marsden-1999} states that with fixed angular momentum 
\Eqref{L} and fixed number of particles \Eqref{Q}, the minimum of $F$ is also a critical 
point of
\begin{equation}
F_\lambda=F +  \lambda_z ( L_z - l_z)  + \lambda_N  (N-1) \,.
\la{Fl} 
\end{equation}
Here $\lambda_z$, $\lambda_N$ are the Lagrange multipliers that enforce the values 
$L_z=l_z$ and $N=1$, respectively. The critical points ($\psi_{cr}, \lambda_z^{cr}, 
\lambda_N^{cr}$)  of $F_\lambda$ \Eqref{Fl} obey 
\begin{align}
-\frac{1}{2}\nabla^2 \psi &+ \frac{|\x|^2}{2} \psi + g |\psi|^2 \psi 
= -  \lambda_N \psi +  i \lambda_z  \ez\cdot\x \wedge  \Grad \psi \, \nonumber\\
&~\text{together with}~~L_z=l_z\,~~\text{and}\,~ N=1\,.
\la{Flambda}
\end{align}
Let $\psi_{min}(\x)$ be a solution of \Eqref{Flambda} that is also a minimum of \Eqref{H}, 
and let $\lambda^{min}_{N}$ and $\lambda^{min}_{z}$ denote the corresponding solutions for 
the Lagrange multipliers.  Whenever the solutions for the Lagrange multipliers do not 
vanish, $\psi_{min}(\x)$ cannot be a critical point of the free energy $F$ in \Eqref{H}.
Instead, it generates a time dependent solution $\psi(\x,t)$ of \Eqref{GPeq} which, 
from \Eqref{Flambda}, with the initial condition $\psi (\x,t=0)=\psi_{min}(\x)$ obeys 
the {\it linear} time evolution  
\begin{equation}
 i \partial_t \psi = -\lambda^{min}_{N}\psi  
			 		+ i\lambda^{min}_{z} \ez\cdot\x \wedge\Grad\psi  \,, \la{Heq1} 
\end{equation}
where both $\lambda^{min}_{N}$ and $\lambda^{min}_{z}$ are time independent 
\cite{Alekseev-2020}.

Whenever either of the Lagrange multipliers is non-vanishing, the minimum energy solution 
is time dependent. But the time evolution amounts to a simple phase multiplication only 
when $\psi_{min}(\x)$ is an eigenstate of the angular momentum operator; below we show that 
this can only occur for $l_z=1$, for all other values of $l_z\neq0$ the minimum energy 
solution always describes a precessing vortex configuration.

The minimum energy wave function $\psi_{min}(\x)$ spontaneously breaks both Noether 
symmetries. The time evolution equation \Eqref{Heq1} describes a symmetry 
transformation of $\psi_{min}(\x)$, that is generated by a definite linear combination 
of the two conserved Noether charges:
\begin{equation}
 i \partial_t \psi =  -\lambda^{min}_{N} \frac{\delta N}{\delta \psi^\star} - 
 \lambda^{min}_{z}  \frac{\delta L_z }{\delta \psi^\star}   
\equiv \  -\lambda^{min}_{N} \{ N , \psi\} -  \lambda^{min}_{z}  
\{ L_z, \psi\} 
 \, ,
 \end{equation}
where $\{ \ , \ \}$ is the Poisson bracket.
This is an example of spontaneous symmetry breaking, but now the symmetry breaking minimum
energy configuration is time dependent \cite{Alekseev-2020}.

\section{Numerical simulations}

To search for minimum energy solutions of the time dependent Gross-Pitaevskii equation, 
we numerically minimize the free energy \Eqref{H} subject to the conditions $L_z=l_z$ 
and $N=1$. The problem is discretized within a finite-element framework \cite{Hecht-2012}, 
and the constrained optimization problem is then solved using the Augmented Lagrangian 
Method (see details of the numerical methods in Appendix~\ref{Sec:Numerics}). 
Minimal energy states for angular momentum values $l_z \in (0,2]$ with $N=1$, and for 
three representative values of the interaction coupling $g=5$, $g=100$ and $g=400$ are 
displayed in figure \ref{Fig:Lagrange-multipliers}; see also 
additional animations in Supplementary materials \cite{supp}.

%
%
%
\begin{figure}[!htb]
\hbox to \linewidth{ \hss
\includegraphics[width=0.9\linewidth]{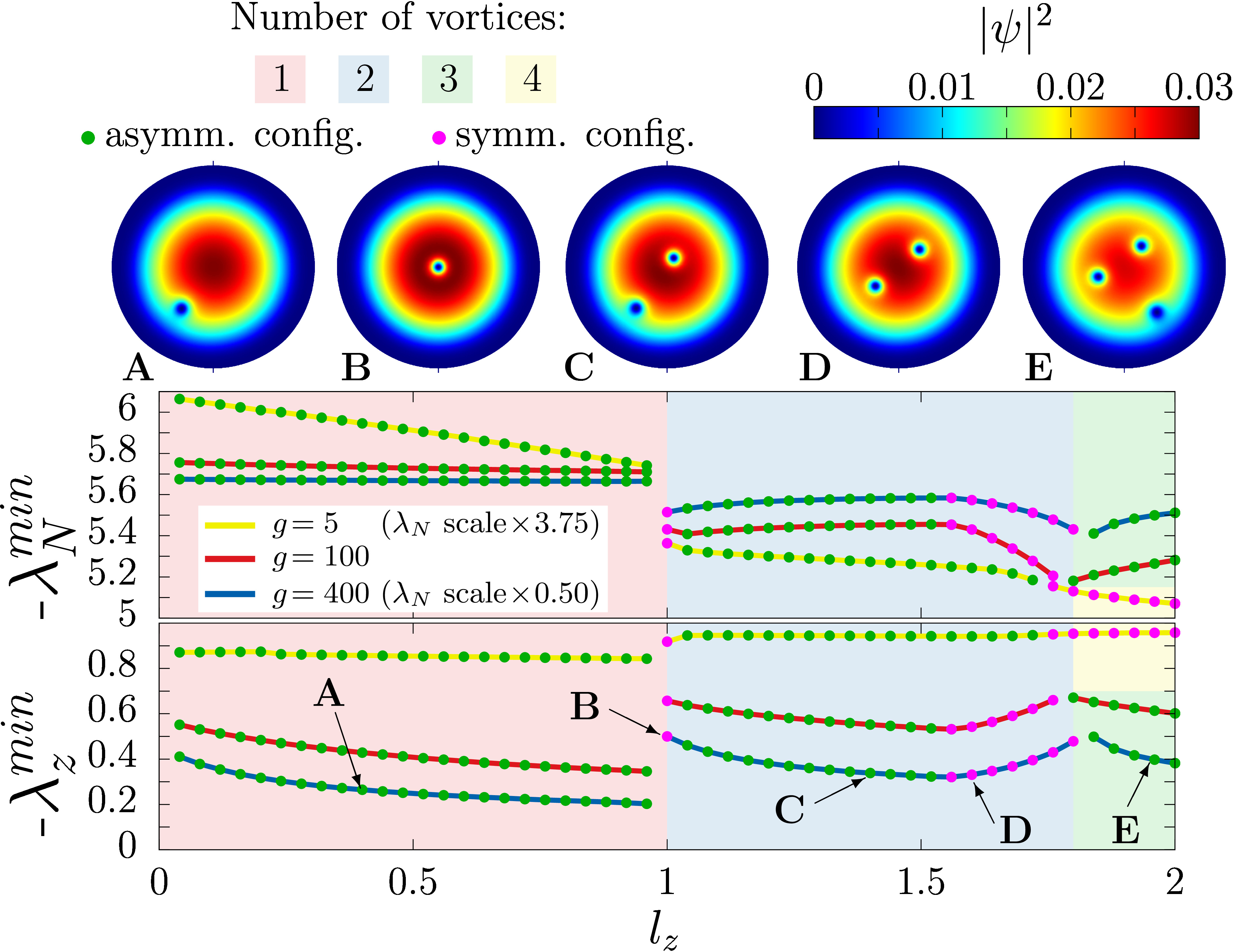}
\hss}
\caption{ 
Minimum energy states for angular momenta $l_z \in (0,2]$. The bottom row shows the 
dependence of the multipliers $\lambda^{min}_N$ and $\lambda^{min}_z$ on the angular 
momentum $l_z$ for $g=5$, $100$ and $400$. The panels on the top row display the density 
of the condensate $|\psi|^2$ for qualitatively different solutions with $g=400$, obtained 
for different values of $l_z$ (zoomed to relevant data while the actual numerical domain 
is larger). At the vortex core the density $|\psi|^2$ vanishes and the phase circulation 
is $2\pi$. The five regimes  A -- E are detailed in the text.  
}
\label{Fig:Lagrange-multipliers}
\end{figure}
%
%
For non-vanishing values of the angular momentum $0<l_z<1$ the minimum energy configuration 
$\psi_{min}(\x)$ is an eccentric vortex that precesses around the trap center (see regime A). 
As $l_z$ increases the vortex core approaches the trap center. This kind of eccentric 
precessing vortices have been previously proposed theoretically in \cite{Kavoulakis-2000}. 
They appear very similar to the precessing vortices that have been observed experimentally 
\cite{Anderson-2000}.

As shown in regime B, when $l_z=1$, the core position coincides with the center of the 
trap, and the Lagrange multipliers feature a discontinuity. At this value of $l_z$, 
our solution coincides with the single vortex solution that has been described extensively 
in the literature \cite{Pethick-2008,Pitaevskii-2003,Lieb-2006,Fetter-2009,Bao-2013,
Aftalion-2001,Seiringer-2002}.
When $l_z$ becomes larger than one, a second vortex appears, and there are two 
qualitatively different two-vortex configurations. As $l_z$ increases, these are 
sequentially asymmetric (regime C) and then symmetric (regime D) with respect to the 
trap center. Because it depends on $g$, there is no universal value of $l_z$ that 
separates these two regimes.
At higher values of angular momentum, $l_z \approx1.8$ with the exact value depending 
on $g$, additional vortices start entering the condensate; this is the regime E in 
figure~\ref{Fig:Lagrange-multipliers}. Remark that for $g=100$ and $g=400$ a third vortex 
moves towards the trap center as $l_z$ increases, while for $g=5$ a pair of vortices 
enters. Further increase of the angular momentum introduces additional vortices in the 
condensate (not shown).

The case $l_z=1$, labelled by B in figure~\ref{Fig:Lagrange-multipliers}, is a special case. 
There, the core of the vortex coincides with the center of the trap. The corresponding 
$\psi_{min}(\x)$ is an angular momentum eigenstate with eigenvalue $l_z =1$, and the 
solution of the time dependent Gross-Pitaevskii equation \Eqref{Heq1} is merely an 
overall phase rotation with no change in the core position. However, for generic $l_z$ 
the minimum energy configuration $\psi_{min}(\x)$ is not an eigenstate of the angular 
momentum. 

We  have simulated the time evolution \Eqref{GPeq} of the minimum energy configuration 
$\psi_{min}(\x)$ using a Crank-Nicolson algorithm. Simulation details together with 
animations of the time evolution can be found in the Appendix~\ref{Sec:Numerics}.
For all values of $l_z\neq1$ the vortex configuration precesses uniformly around the 
trap center. This uniform rotation is a consequence of the equation \Eqref{Heq1} which 
determines the time evolution of the minimal energy configurations, in terms of the 
two conserved charges. 

\section{Exchange of vortices}

\begin{figure}[!tb]
\hbox to \linewidth{ \hss
\includegraphics[width=0.9\linewidth]{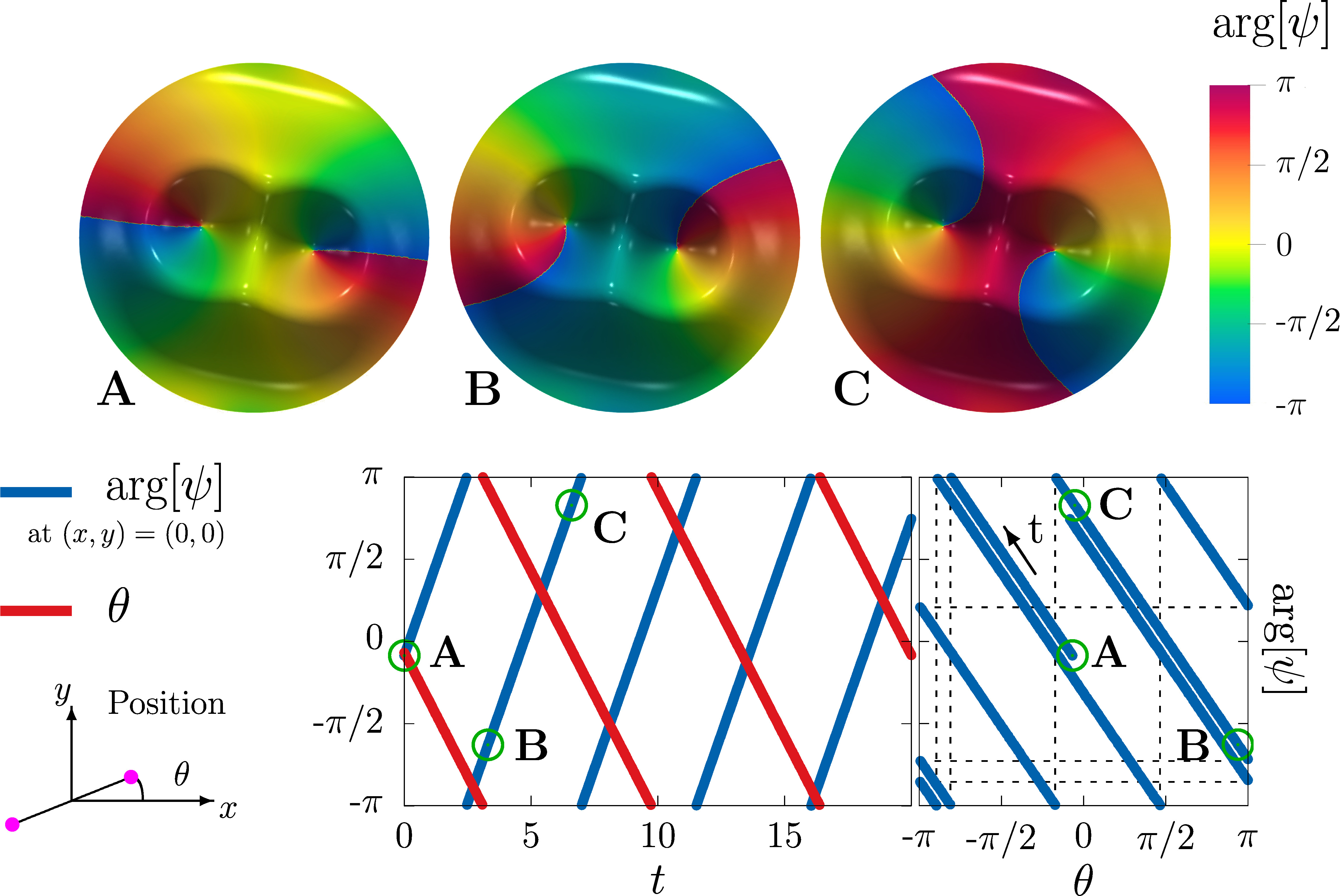}
\hss}
\caption{ 
The evolution of a minimum energy configuration $\psi_{min}(\x)$, in the case of a 
symmetric vortex pair (with $g=5$) in region D of figure \ref{Fig:Lagrange-multipliers}.
The bottom panels show the evolution of $\theta$, the rotation angle of the pair, and 
of the value of the phase $\arg[\psi]$ measured at the center of the trap. 
The three panels A--C on the top row are snapshots displaying the phase $\arg[\psi]$ 
after rotations $\theta=\theta_0$, $\theta=\theta_0+\pi$ and $\theta=\theta_0+2\pi$ 
respectively.
Since the Lagrange multipliers $\lambda_N^{min}$ and $\lambda_z^{min}$ are not 
commensurate, $\arg[\psi](t)$ and $\theta(t)$ feature different periodicities. 
As a result after $\pi$ rotations of the pair the phase profile is not simply 
an overall phase multiplication. 
}
\label{Fig:Time-evolution}
\end{figure}

Finally, an interesting phenomenon occurs for the time evolution of a symmetric pair 
of identical vortices (corresponding to the regime D of figure 
\ref{Fig:Lagrange-multipliers}). The time evolution of such a vortex pair, with $g=5$, 
is displayed in figure~\ref{Fig:Time-evolution}.
The vortex pair rigidly rotates around the trap center at a constant distance, with 
an angular velocity specified by $\lambda^{min}_{z}  \equiv -\Omega$. In addition the 
phase of the wave function evolves with a rate that depends on $\lambda^{min}_{N}$. 
This coincides with the rate of change in the phase $\arg[\psi](t)$, measured at the 
center of the trap. Similarly, the rate of change of $\theta(t)$ that measures the 
rotation of the pair, equals $\lambda^{min}_{z}$ . 
Since the two Lagrange multipliers are in general not commensurate, it follows that 
when the time evolution exchanges the position of two identical vortices, the change in 
the phase of the wave function is not by an integer multiple of $\pi$.

%
%
%

The quantum mechanical exchange of two identical bodies, with ensuing phase change, 
is familiar from description of anyons \cite{Leinaas-1977,Wilczek-1982a,Wilczek-1982}. 
The similarity to that scenario can be elaborated further, by following \cite{Wilczek-1982a} 
to rewrite the time evolution \Eqref{Heq1} as follows:
\begin{equation}
 i \partial_t \psi = \Omega\, \vL^{cov}_z  \psi \ \equiv \
 \Omega (  \ez\cdot\vL +  \phi \, \ez \cdot \x \wedge  \A )\psi \,.
 \la{anyon}
\end{equation}
Here $\vL = - i  \x \wedge \Grad$ is the canonical angular momentum, the parameter 
$\phi = \lambda^{min}_{N} / \lambda^{min}_{z}$, and the flux of $\A = \Grad\tan^{-1}x/y$ 
equals $2\pi$ around any counter-clockwise path that encircles the $z$-axis once. 
Notably, in the case of two identical vortices $\x$ coincides with their relative 
coordinate in a center-of-mass frame. Furthermore, the equation \Eqref{anyon} is akin a 
time dependent {\it linear} Schr\"odinger equation that describes the evolution of a 
charged point particle, moving on the ($x,y$) plane under the influence of a unit 
strength magnetic vortex filament along the $z$-axis. The corresponding Hamiltonian 
comprises the $z$-component of the covariant angular momentum $\vL^{cov}_z$. We note that 
when $\phi\not=0$, the rotations around $z$-axis, {\it i.e.} the changes in the polar 
angle $\theta$, are generated by the covariant angular momentum.

To find the solutions of \Eqref{anyon} we first remove $\A$ by phase multiplication, 
and then proceed with separation of variables in terms of polar coordinates ($r,\theta$).
\begin{equation}
\psi(\mathbf x,t) \ =  \  e^{-iEt} \mathcal X(r,\theta) \, 
e^{ - i \phi \!  \int\limits_{\x_0}^{\x} \A \cdot d\bs{\ell} }\,,
\end{equation}
The single valuedness of the wave function $\psi(\x,t)$ implies that 
$\mathcal X (r,\theta)$ is generally not single valued as it obeys
\begin{equation}
\vL_z  \mathcal X_n = (n + \phi) \mathcal X_n \,.
\end{equation}
The solution of the time evolution \Eqref{anyon} is thus
\begin{equation}
\psi(\x,t) = e^{ - i \phi \!  \int\limits_{\x_0}^{\x} \A\cdot d\bs{\ell} }
\sum_n f_n(r) e^{ - i (n+\phi) ( \Omega t-\theta ) } \,.
\la{solution}
\end{equation}
Here the $f_n(r)$ form an orthogonal set of normalizable, real valued functions, they are
selected so that \Eqref{solution} describes the energy minimizer $\psi_{min}(\x)$ of  
\Eqref{Flambda}, \Eqref{GPeq} at $t=0$.
%

\section{Summary and outlook}

In summary, we have investigated the time dependent two-dimensional Gross-Pitaevskii 
equation, that models dynamics 
in an axially symmetric harmonic trap. The equation supports two Noether 
charges that correspond to the number of atoms and to the axial component of the 
macroscopic angular momentum, respectively.
We have used the methodology of constrained optimization to construct solutions that 
minimize the Gross-Pitaevskii free energy, with a fixed number of atoms and 
as function of the macroscopic angular momentum 
to which we have assigned arbitrary numerical values. For generic angular momentum values, 
a minimum energy configuration always exists and it describes eccentric vortices that 
precess uniformly around the center of the trap. In particular, no time independent 
solution can exist in a non-rotating cylindrically symmetric trap.
Moreover, whenever the solution describes two identical vortices, the time evolution 
exchanges their positions but the phase of their common wave function evolves in a 
nontrivial fashion. 

\vskip 0.2cm
We foresee that the vortex precession in a non-rotating condensate that we have described,  
can be realized and observed in a variety of laboratory experiments; the precessing 
vortices already reported in \cite{Anderson-2000} appear to be very similar to the 
vortex precession that we have simulated. There are several conceivable experimental 
setups, where vortices can form in a non-rotating condensate. 
These include vortex formation by optical photon beams that carry a non-vanishing angular 
momentum, with appropriate topologies \cite{Marzlin-1997,Bolda-1998,Dum-1998,Leanhardt-2002}.
Additional experimental set-ups can be built on thermal quenching in combination with 
the Kibble-Zurek mechanism \cite{Kibble-1976,Zurek-1985,Zurek-1996,Anglin-1999}. This 
mechanism describes how topological defects form during rapid symmetry breaking 
transitions. There are already several laboratory examples that are relevant to the 
scenario we have studied, including neutron irradiation in the case of superfluid $^3$He 
\cite{Ruutu-1996,Eltsov-2000}, evaporative cooling in the case of ultracold $^87$Rb 
atoms \cite{Weiler-2008,Freilich-2010}.

\vskip 0.4cm
\begin{acknowledgments}
AJN thanks C. Pethick  and K. Sacha for discussions. 
We thank Rima El Kosseifi for discussions. 
The work by DJ and AJN is supported by the Carl Trygger Foundation Grant CTS 18:276
and by the Swedish Research Council under Contract No. 2018-04411. DJ and AJN also 
acknowledge collaboration under COST Action CA17139. The research by AJN was also 
partially supported by Grant No. 0657-2020-0015 of the Ministry of Science and 
Higher Education of Russia 
The computations were performed on resources provided by the Swedish National 
Infrastructure for Computing (SNIC) at National Supercomputer Center at Link\"{o}ping, 
Sweden. 

\end{acknowledgments}

\section*{Appendices}

\appendix

\section{Nondimensionalization of the Gross-Pitaevskii equation}
\label{Sec:Dimensionless}

In the main body of the paper, the dynamics of a Bose-Einstein ultracold atoms is 
described by the \emph{dimensionless} Gross-Pitaevskii equation. Here, we derive this 
dimensionless form from the original equation (see e.g. \cite{Bao-2013}). In the 
ultracold dilute regime, the Bose-Einstein condensate state of $N_a$ identical atoms 
is described by the macroscopic wave function $\Psi$, whose dynamics obey 
\SubAlign{Eq:Dim:dimensionful}{
i\hbar\partial_t\Psi&=\left[-\frac{\hbar^2}{2m}\nabla^2 +V(\x)
+\frac{N_a}{N}g_0|\Psi|^2\right]\Psi \,, 	\\
\text{where}~~&N=\int|\Psi|^2
\,, ~~\text{and}~~g_0=\frac{4\pi\hbar^2a_s}{m} \,.
}
Here, $\nabla^2\equiv\Div\Grad$ is the Laplace operator, $m$ is the mass of the atoms, 
$\hbar$ is the reduced Planck's constant, and $a_s$ is $s$-wave scattering length of 
the atoms. Note that for practical purposes, we introduce $N$ to be the norm of the 
condensate, while $N_a$ is the actual number of atoms in the condensate. 
The harmonic oscillator trap potential reads as 
\Equation{Eq:Dim:trap}{
V(\x)=\frac{m}{2}\big(\omega^2_x x^2+\omega^2_y y^2+\omega^2_z z^2\big)
 \,,
}
where $\omega_x\leq\omega_y\leq\omega_z$ are the trapping frequencies in the different 
spatial directions. The dimensionless units are obtained by the following rescaling 
\Equation{Eq:Dim:rescaling}{
 t=t_s \tilde{t} \,,~~ \x=x_s \tilde{x} \,,~~\Psi=x_s^{-3/2} \psi 
\,,~~\text{where}~~ t_s=\frac{1}{\omega_x}
\,,~~\text{and}~~ x_s=\sqrt{\frac{\hbar}{m\omega_x}} \,.
}
Hence, after dropping the $\tilde{\,}$, the dimensionless Gross-Pitaevskii equation 
is

\SubAlign{Eq:Dim:dimensionless}{
i\partial_t\psi&=\left[-\frac{1}{2}\nabla^2+V(\x)+g|\psi|^2\right]\psi \,, \\
~\text{where}~~V(\x)=\frac{1}{2}\sum_{\alpha=x,y,z}&
			\left(\frac{\omega_\alpha}{\omega_x}\right)^2x^2_\alpha
\,, ~N=\!\int|\psi|^2
\,, ~\text{and}~~g=\frac{4\pi a_sN_a}{x_sN} 
\,.
}

The dimensionless time dependent Gross-Pitaevskii equation \Eqref{Eq:Dim:dimensionless}, 
is obtained for a three dimensional harmonic trap. In the main body of the paper, we 
consider the case of a two Bose-Einstein condensate. The two dimensional problem fairly 
approximates a disk-shaped condensate with small height in $z$-direction, for the trap 
frequencies $\omega_x\approx\omega_y$ and $\omega_z\gg\omega_x$. This results in a further 
rescaling of the coupling constant $g_{2d}= g_{3d}\sqrt{\frac{\omega_z}{2\pi\omega_x}}$. 
For detailed derivation see, e.g., \cite{Bao-2013}.

The dimensionless coupling $g$ is thus the parameter of the accounts for the interactions 
between atoms. In the numerical investigations we choose values $g$ that are representative 
of a Bose-Einstein condensate of ultracold alkali atoms. The Table \ref{Tab:values}
shows typical experimental values of trapped ultracold atoms, and the corresponding 
estimates for the length scales and the dimensionless coupling $g$.

\begin{table}[!ht]
\centering
\begin{tabular}{|c||c|c|c|c|}
\hline 
\rule[-1ex]{0pt}{2.5ex}  Atoms & $m\ (kg)$ 
		& $a_s\ (m)$ & $\omega_x\ (rad.s^{-1})$ & $N_a$  \\ 
\hline \hline 
\rule[-1ex]{0pt}{2.5ex} $^{87}Rb$ \cite{Anderson-1995} & $1.44\times10^{-25}$ 
		& $5.1\times10^{-9}$ & 20 & $10^{2}\sim10^{7}$ 	\\ 
\hline 
\rule[-1ex]{0pt}{2.5ex} $^{23}Na$ \cite{Davis-1995}& $3.817\times10^{-26}$ 
		& $2.75\times10^{-9}$ & $2\pi\times200$ & $\times10^{4}\sim5\times10^{5}$ \\ 
\hline 
\end{tabular} 
\vspace{0.01cm}
\hspace{-4cm}
\begin{tabular}{|c||c|c|c|}
\hline 
\rule[-1ex]{0pt}{2.5ex}  Atoms 
		& $x_s\ (m)$ & $\frac{4\pi a_s}{x_s}$ & $g\ (\times N=1)$ \\ 
\hline \hline 
\rule[-1ex]{0pt}{2.5ex} $^{87}Rb$ \cite{Anderson-1995} 
		& $6.0\times10^{-6}$ & $1.1\times10^{-2}$ & $1\sim10^{5}$	\\ 
\hline 
\rule[-1ex]{0pt}{2.5ex} $^{23}Na$ \cite{Davis-1995}
		& $1.5\times10^{-6}$ & $2.3\times10^{-2}$ & $10^{2}\sim10^{4}$\\ 
\hline 
\end{tabular} 
\caption{Typical physical values for various atomic Bose-Einstein condensates, 
and the resulting scale and dimensionless coupling. The reduced Planck constant is 
$\hbar=1.05\times10^{-34}J.s$.
}
\label{Tab:values}
\end{table}

\section{Constrained minimization with fixed angular momentum  \texorpdfstring{\\}{}  
		versus minimization at constant angular velocity }
 \label{Sec:Comparison}

In the main body of the paper, we emphasize that there is a difference between the 
common set-up that describes vortices in a rotating trap in terms of a free energy 
expressed in the co-rotating frame \cite{Butts-1999,Aftalion-2001,Seiringer-2002,
Pitaevskii-2003,Lieb-2006,Pethick-2008,Fetter-2009,Bao-2013}, and the present set-up 
where vortices occur in a non-rotating trap; the present set-up was first addressed 
in \cite{Mottelson-1999,Kavoulakis-2000}. Here we clarify the difference, by a direct 
comparison of the two cases:

In the case of a rotating trap, vortices are created by rotating the trap, with a 
constant angular velocity $\Omega$. The free energy $F^\Omega$ is expressed in the 
co-rotating frame, and the vortices are the critical points of the free energy.  
The vortex structures can then be investigated as a function of the externally applied 
angular velocity $\Omega$.
In contrast, in our case the trap does not rotate. Instead, we use the fact that the 
angular momentum $L_z$ of the condensate is a conserved quantity, and the condensate 
carries a corresponding angular momentum value $L_z=l_z$.  
The ensuing minimal free energy configuration is a solution of a \underline{constrained} 
optimization problem that minimizes the free energy $F$, subject to the condition 
$L_z=l_z$. The two free energies are related as follows:
\Equation{Eq:GPenergy:Rotating}{
F^\Omega = F - \Omega L_z 
\,,~~~~\text{where}~~
L_z  \ \equiv \ \int d^2x\, \psi^\star\, \ez\cdot(- i\hbar\,\x\wedge \Grad)\psi 
\,.
}
Due to the constraint, the free energy minimizer is in general not a critical point 
of $F$. But it can be located using the Lagrange multiplier theorem \cite{Marsden-1999},  
in the framework of constrained optimization \cite{Fletcher-1987,Nocedal-1999}. 
Whenever the minimizer of the free energy $F$ is not a critical point, it describes
a solution of a \underline{time dependent} Gross-Pitaevskii equation.

In addition, in both cases the norm of the condensate  wave function  ({\it i.e.} the 
particle number) needs to be accounted for. Here the goal is to clarify the difference 
between the rotating and non-rotating cases. Therefore, in our comparison we treat the 
constraint $N=1$ the same way, both when searching for the minimum of $F^\Omega$ and $F$, 
by using the Lagrange multiplier theorem. For the problem we consider, the Lagrange 
multiplier theorem \cite{Marsden-1999} states that with fixed angular momentum and fixed 
number of particles $N$, the minimum of $F$ is also a critical point of 
\Equation{Eq:LagMult:1a}{
F_\lambda\equiv F +  \lambda_z ( L_z - l_z)  + \lambda_N  (N-1) \,.
}
Here $\lambda_z$, $\lambda_N$ are the Lagrange multipliers that enforce the values 
$L_z=l_z$ and $N=1$ respectively. On the other hand, for a rotating condensate, 
the minimum of $F^\Omega$ is a critical point of the following free energy
\Equation{Eq:LagMult:1b}{
F_\lambda^\Omega\equiv F^\Omega  + \lambda_N  (N-1) 
=F +  \Omega  L_z   + \lambda_N  (N-1) \,, 
}
where $\lambda_N$ is the (only) Lagrange multiplier that enforces $N=1$. Note that the 
present Lagrange multiplier $\lambda_N$ is commonly treated as a chemical potential 
that is an externally controlled parameter, in par with $\Omega$.
This is not the case here, as $\lambda_N$ has to be adjusted to satisfy the constraint.
The critical points of $F_\lambda$ and $F_\lambda^\Omega$ obey respectively
\SubAlign{Eq:LagMult:2}{
F_\lambda: ~~-\frac{1}{2}\nabla^2 \psi &+ \frac{|\x|^2}{2} \psi + g |\psi|^2 \psi 
= - \lambda_N \psi + i \lambda_z  \ez\cdot\x \wedge  \Grad \psi \,,
\label{Eq:LagMult:2a} \\
&\text{and the conditions}~L_z=l_z~\text{and}~N=1 
\,,\nonumber\\
F_\lambda^\Omega: ~~-\frac{1}{2}\nabla^2 \psi &+ \frac{|\x|^2}{2} \psi + g |\psi|^2 \psi 
- i \Omega\,\,\ez\cdot\x \wedge  \Grad \psi = - \lambda_N \psi \,, 
\label{Eq:LagMult:2b}\\
&\text{and the condition}~N=1 \,. \nonumber
}
%
Note that in a numerical simulation $\Omega$ is an input parameter that is fixed during 
the simulation, while $\lambda_z$ is adjusted sequentially, to satisfy the constraint 
(according to the Augmented Lagrangian Method, see details in Sec.~\ref{Sec:ALM}). 
The differences between the two approaches are summarized at the end of this section 
in Table~\ref{Tab:Summary}.

\begin{table}[!ht]
\centering
\begin{tabular}{|c||c|c|c|c|c||}
\hline 
\rule[-1ex]{0pt}{2.5ex}   Critical  &  Obey the equation 
&  Satisfy &  Control    &  Mult. 
&  Output\\ 
\rule[-1ex]{0pt}{2.5ex}   pts. of &  Obey the equation 
&   &  params.   &   
&  \\ 
\hline \hline 
\rule[-1ex]{0pt}{2.5ex}  $F_\lambda$ &  
$-\frac{1}{2}\nabla^2 \psi + \frac{|\x|^2}{2} \psi + g |\psi|^2 \psi =$
& $L_z\!=\Red{l_z}$  & $g$, $\Red{l_z}$ 
&  $\Purple{\lambda_N}$,
& $\Olive{\lambda_N^{min}}$, $\Olive{\lambda_z^{min}}$		\\ 
&\hspace{0.5cm}$-\Purple{\lambda_N}\psi + i\Purple{\lambda_z}\,\,
\ez\cdot\x\wedge\Grad\psi$ 
~& and $N=1$&&  $\Purple{\lambda_z}$ & $\Olive{\psi_{min}}$\\
\hline 
\rule[-1ex]{0pt}{2.5ex}  $F_\lambda^\Omega$  &  
$-\frac{1}{2}\nabla^2 \psi + \frac{|\x|^2}{2} \psi + g |\psi|^2 \psi $
& $N=1$ &  $g$, $\Red{\Omega}$ & $\Purple{\lambda_N}$
& $\Olive{\lambda_N^{min}}$, 		\\  
&\hspace{0.5cm}$- i\Red{\Omega}\,\,\ez\cdot\x\wedge\Grad\psi=-\Purple{\lambda_N}\psi$ 
~&&& &$\Olive{\psi_{min}}$\\
\hline 
\end{tabular} 
\caption{
Summary of the differences:  The \Red{red} font shows the parameters that are given as 
input to the corresponding minimization problem. The \Purple{purple} font highlights 
the Lagrange multipliers. In a numerical simulation, these are sequentially adjusted 
according the Augmented Lagrangian Algorithm, until the constraints are satisfied (and 
the minima of $F_\lambda$ are found). The \Olive{olive} color fonts show the output 
quantities obtained in the numerical simulations. 
}
\label{Tab:Summary}
\end{table}

\begin{figure}[!htb]
\hbox to \linewidth{ \hss
\includegraphics[width=0.8\linewidth]{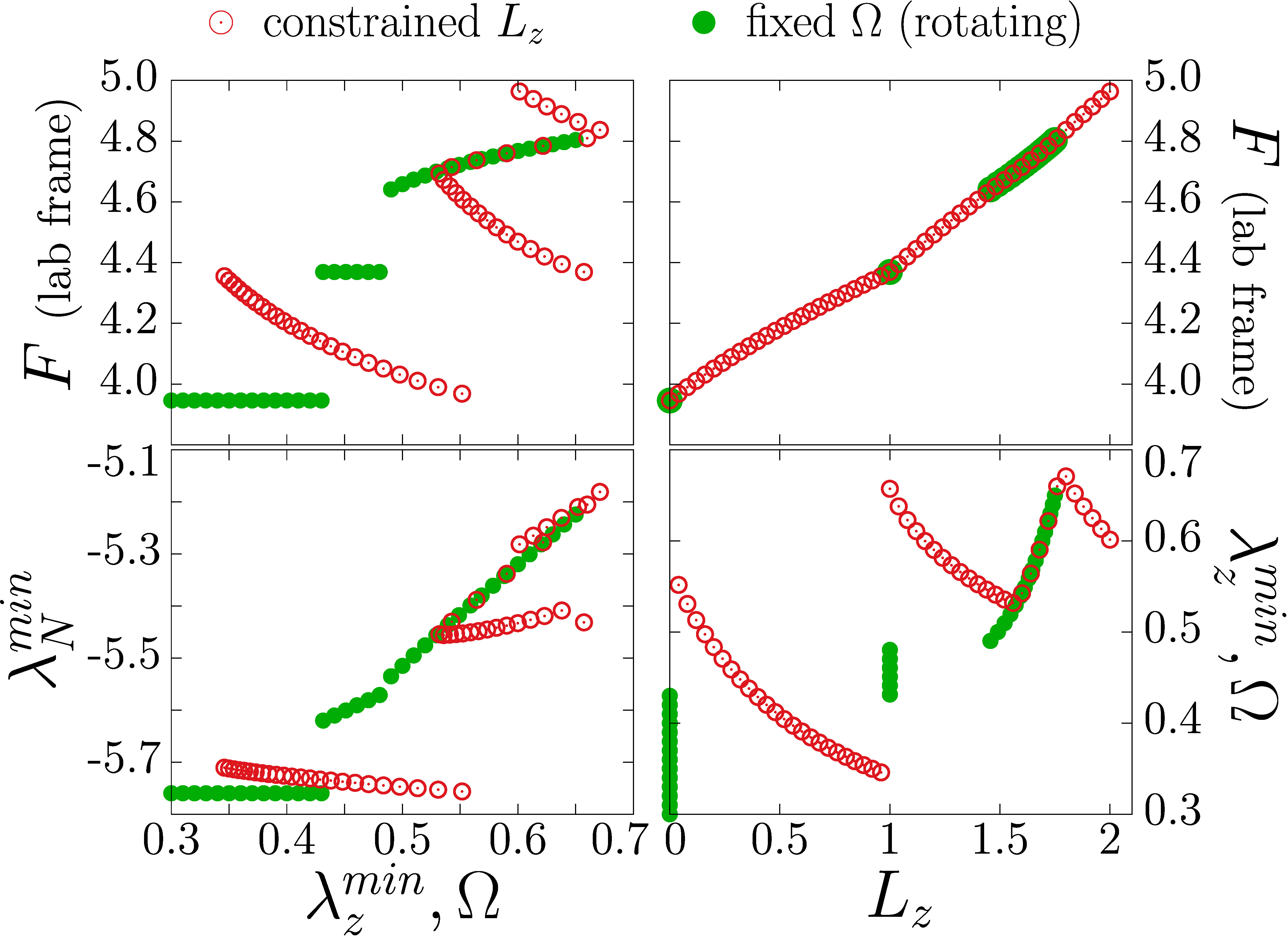}
\hss}
\caption{ 
Detailed comparison  of the solutions to the two equations (\ref{Eq:LagMult:2}) 
for $g=100$.
The panels on the top row display the energy, in a non-rotating frame, as a function 
of \emph{angular velocity} $\Omega$  (left) and value of \emph{angular momentum} $L_z$ 
(right). 
The empty (red) circles denote the results from the constrained minimization at given 
value of the angular momentum $L_z$. The filled (green) circles show the results
from the minimization of a rotated condensate at given angular velocity $\Omega$.
}
\label{FigApp:Compare}
\end{figure}

Comparison of the minimal energy solutions of the two equations \Eqref{Eq:LagMult:2}, 
can be seen in figure~\ref{FigApp:Compare}. The results are indeed quite different.
In particular, the free energy (evaluated in a non-rotating laboratory frame) that 
is displayed on the panels on the top row of figure~\ref{FigApp:Compare}, show that 
the minimization with fixed angular frequency $\Omega$ cannot support solutions 
with arbitrary values of angular momentum. Indeed, there are excluded ranges  
\begin{equation}
L_z\in[0,1[~\cup~]1,1.457[~\cup~]1.75,2.68[\cdots\,,
\label{ranges}
\end{equation} 
where there are no \emph{rotation-induced} stable vortex states (see the green dots 
on the top right panel). This is in agreement with the known results for rotating 
Bose-Einstein condensates; compare for example the green dots in the top left panel 
of figure \ref{FigApp:Compare}, {\it e.g.} with the results shown in figure~2 of 
\cite{Butts-1999}.
In contrast, in the present case, there cannot be any excluded values of the angular 
momentum in the equation \Eqref{Eq:LagMult:2a} (see the red dots on the top right panel). 
Those vortex configurations with angular momentum in the excluded ranges \Eqref{ranges} 
describe eccentric asymmetric vortex configurations that precess around the trap center. 

Observe that in the case of equation \Eqref{Eq:LagMult:2a}, there is multi-valuedness 
in the free energy (red dots), as a function of the Lagrange multiplier $\lambda_z$; 
see for example the top left panel at $\lambda_z^{min}\approx0.55$  which displays 3 
different states at the same value of $\lambda_z$. These are a single eccentric vortex, 
an asymmetric vortex pair, and a symmetric vortex pair for increasing free energy,
respectively. These configurations have both a different angular momentum value, and 
a different value of the Lagrange multiplier $\lambda_N^{min}$ (see the bottom panels) 
which plays the role of a chemical potential.
The present comparisons, summarized in figure~\ref{FigApp:Compare}, show that the 
two equations \Eqref{Eq:LagMult:2a} and \Eqref{Eq:LagMult:2b} do indeed describe two 
quite different physical scenarios.

\section{Details of the numerical methods}
\label{Sec:Numerics}

In the numerical investigations in the main body of the paper, we use Finite-Element 
Methods (FEM) (see e.g. \cite{Hutton-2003,Reddy-2005}) both for the minimization and 
for the time-evolution. In practice we use the finite-element framework provided by 
the FreeFEM library \cite{Hecht-2012}.
Within the finite-element framework, the constrained minimization is addressed using 
an Augmented Lagrangian Method together within a non-linear conjugate gradient 
algorithm. The time-dependent Gross-Pitaevskii equation, is integrated using 
a Crank-Nicolson algorithm with a forward extrapolation of the nonlinear term.

\subsection{Finite-element formulation}

We consider the domain $\Omega$ which a bounded open subset of $\Real^2$ and denote 
$\partial\Omega$ its boundary. $H(\Omega)$ stands for the Hilbert space, such that a 
function belonging to $H(\Omega)$, and its weak derivatives have a finite $L^2$-norm. 
Furthermore, $\mathcal{H}(\Omega)=\lbrace u+iv\,|\,u,v\in H(\Omega)\rbrace$ denotes the 
Hilbert spaces of complex-valued functions. The Hilbert spaces of real- and complex-valued 
functions are equipped with the inner products $\langle\cdot,\cdot\rangle$, defined as:
\Equation{Eq:Inner}{
\ScalarProd{u}{v}=\int_\Omega uv  \,,~\text{for}~u,v\in H({\Omega}) 
\,,~~~~\text{and}~~~~
\ScalarProd{u}{v}=\int_\Omega u^\star v\,,~\text{for}~u,v\in\mathcal{H}({\Omega}) \,.
}
The spatial domain $\Omega$ is discretized as a mesh of triangles using for the 
Delaunay-Voronoi algorithm, and the regular partition $\mathcal{T}_h$ of $\Omega$ refers 
to the family of the triangles that compose the mesh. Given a spatial discretization, 
the functions are approximated to belong to a \emph{finite-element space} whose properties 
correspond to the details of the Hilbert spaces to which the functions belong. We define 
${P}^{(2)}_h$ as the $2$-nd order Lagrange finite-element subspace of $H(\Omega)$, and 
correspondingly $\mathcal{P}^{(2)}_h$ for $\mathcal{H}(\Omega)$. Now, the physical 
degrees of freedom can be discretized in their finite element subspaces. And we define 
the finite-element description of the degrees of freedom as 
$\psi\mapsto \psi^{(h)}\in \mathcal{P}^{(2)}_h$. This describes a linear vector space 
of finite dimension, for which a basis can be found. The canonical basis consists of 
the shape functions $\phi_k(\x)$, and thus 
\Equation{Eq:FEM:space}{
V_h(\mathcal{T}_h,\mathrm{P}^{(2)})=\Big\lbrace w(\x)=\sum_{k=1}^M w_k\phi_k(\x)
,\phi_k(\x)\in \mathrm{P}^{(2)}_h
\Big\rbrace\,.
}
Here $M$ is the dimension of $V_h$ (the number of vertices), the $w_k$ are called 
the degrees of freedom of $w$ and M the number of the degrees of freedom.
To summarize, a given function is approximated as its decomposition:
$w(\x)=\sum_{k=1}^M w_k\phi_k(\x)$, on a given basis of shape functions $\phi_k(\x)$ 
of the polynomial functions $\mathrm{P}^{(2)}$ for the triangle $T_{i_k}$. 
The finite element space $V_h(\mathcal{T}_h,\mathrm{P}^{(2)})$ hence denotes the 
space of continuous, piecewise quadratic functions of $x$, $y$ on each triangle of 
$\mathcal{T}_h$.

\subsection{Constrained minimization: Augmented Lagrangian Method}
\label{Sec:ALM}

In the main body of the paper, we aim to minimize the free energy while enforcing 
a set of two conditions. Such problems, referred to as \emph{constrained optimization}
are studied in great details, see for example textbooks \cite{Gill-1981,Fletcher-1987,
Nocedal-1999,Birgin-2008}. Here we describe the numerical algorithms that were used in 
the main body of the paper, to solve the constrained optimization problem. The Augmented 
Lagrangian Method (ALM) used to solve the constrained optimization is based on the 
following. In terms of the original energy functional to be minimized $F$, and the set 
of conditions $C_i$, the augmented Lagrangian $F^\mathrm{aug}$ is defined as 
\Equation{Eq:ALM:Hamiltonian}{
F^\mathrm{aug}:= F + \frac{\mu}{2} \left( \sum_j C_j^2 \right)
	+\sum_j \lambda_jC_j \,.
}
Here $\mu$ is a penalty parameter and $\lambda_j$ are the Lagrange multipliers associated 
with the conditions $C_j$. In the Augmented Lagrangian Method, the augmented Lagrangian 
is minimized, and the penalty $\mu$ is iteratively increased while the multipliers 
$\lambda_j\leftarrow\lambda_j+\mu C_j$ until all conditions are satisfied with a specified 
accuracy. The ALM algorithm has the property to converge in a finite number of iterations.
Our choice for the minimization algorithm within each ALM iteration is a non-linear 
conjugate gradient algorithm \cite{Hestenes-1952,Fletcher-1964,Polak-1969}.

In this work, the minimized functional $F$ is the dimensionless Gross-Pitaevskii free 
energy \cite{Bao-2013} 
\Equation{Eq:GPenergy}{
F=\int d^2x \left\{\frac{1}{2}|\Grad\psi|^2 +V(\x)|\psi|^2+\frac{g}{2}|\psi|^4\right\}\,,
}
where $g$ is the dimensionless coupling that accounts for the interatomic interactions, 
and $V(\x)$ is the trapping potential. In the main body we considered an harmonic 
oscillator trapping potential $V(\x)=|\x|^2/2$. Since the derivations here do not 
depend on the specific form of the potential, we keep in general in the following. 
The two conditions specifying the particle number $N=1$ and the value $l_z$ of the 
angular momentum are
\Equation{Eq:ALM:Constraints}{
C_N=N-1\equiv\int d^2x |\psi|^2-1 ~~~~\text{and}~~~~ C_{z}=L_z-l_z\equiv\int d^2x \, 
\psi^\star\big[\ez\cdot(-i\x\wedge\!\Grad)\big]\psi -l_z   \,.
}
Hence, the variations of the augmented Lagrangian 
with respect to $\psi^\star$ and $\lambda_i$
give the set of equations
\SubAlign{Eq:EL}{\label{Eq:EL:1}
-\frac{1}{2}\nabla^2\psi +\left[V(\x)+g|\psi|^2\right]\psi 
+ {\tilde\lambda}_N\psi
+ {\tilde\lambda}_z(\ez\!\cdot\!\vL)\psi &=0 
\,, ~~~\text{where}~{\tilde\lambda}_i=(\mu C_i+\lambda_i)
\\
C_i&=0\,, ~~~\text{and}~i=N,z
\,,
}
where the angular momentum operator is $\vL=-i\x\wedge\!\Grad$.
The associated weak form, is obtained by multiplying the equation \Eqref{Eq:EL:1} 
by test functions $\psi_w\in\mathcal{H}(\Omega)$, integrating over $\Omega$ and 
integrating by parts the Laplace operator. Alternatively the weak form is calculated 
as the Fr\'echet derivative of the augmented Lagrangian as \Eqref{Eq:ALM:Hamiltonian}. 
In terms of the inner products \Eqref{Eq:Inner}, we find
\Equation{Eq:Minim:1}{
 [\psi_w ]\cdot F^\prime( \psi ) =
\frac{1}{2}\ScalarProd{\Grad\psi_w}{\Grad\psi}
+\ScalarProd{\psi_w}{\left[V(\x)+g|\psi|^2\right]\psi}
+{\tilde\lambda}_N\ScalarProd{\psi_w}{\psi}
+{\tilde\lambda}_z\ScalarProd{\psi_w}{\ez\!\cdot\!\vL\psi}\,.
}
This formulation, hence corresponds to the variations of the Lagrangian 
with respect to all degrees of freedom. It can be seen as the gradient of a function 
to be minimized. To do so we choose a nonlinear conjugate gradient algorithm.

\subsubsection*{Error estimates}

To estimate the quality of the solution, a global error can be derive by multiplying 
\Eqref{Eq:EL:1} by $\psi^\star$ and integrating over the domain $\Omega$ (and again 
integrating by part the Laplace operator). The error is alternatively obtained by 
replacing the test functions $\psi_w$ by $\psi$ in \Eqref{Eq:Minim:1}:
\SubAlign{}{
\int\frac{1}{2}|\Grad\psi|^2 +\left[V(\x)+g|\psi|^2\right]|\psi|^2 +
{\tilde\lambda}_q\int |\psi|^2
+{\tilde\lambda}_z\int \psi^\star(\ez\!\cdot\!\vL)\psi=&0 \\
F+\frac{g}{2}\int|\psi|^4 +
{\tilde\lambda}_N(C_N+1)
+{\tilde\lambda}_z(C_z+l_z)=&0 \\
1+\frac{\sum_i{\tilde\lambda}_i(C_i-l_i)}{F+\frac{g}{2}\int|\psi|^4}=&~err.
}
In pratice during our constrained minimization simulations, we obtain the typical 
values for the relative error to be around $10^{-5}$.

\begin{figure}[!htb]
\hbox to \linewidth{ \hss
\includegraphics[width=0.5\linewidth]{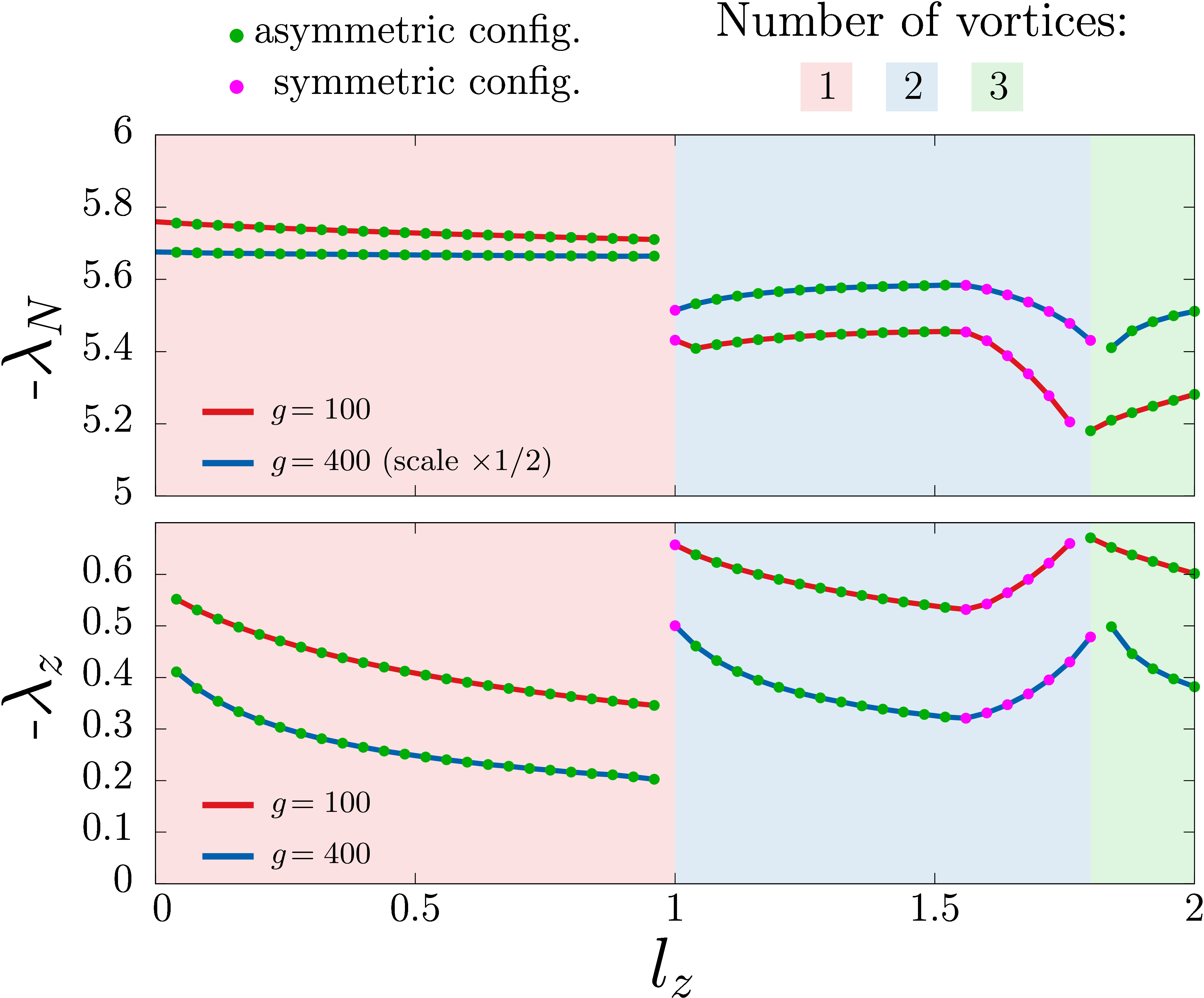}
\includegraphics[width=0.5\linewidth]{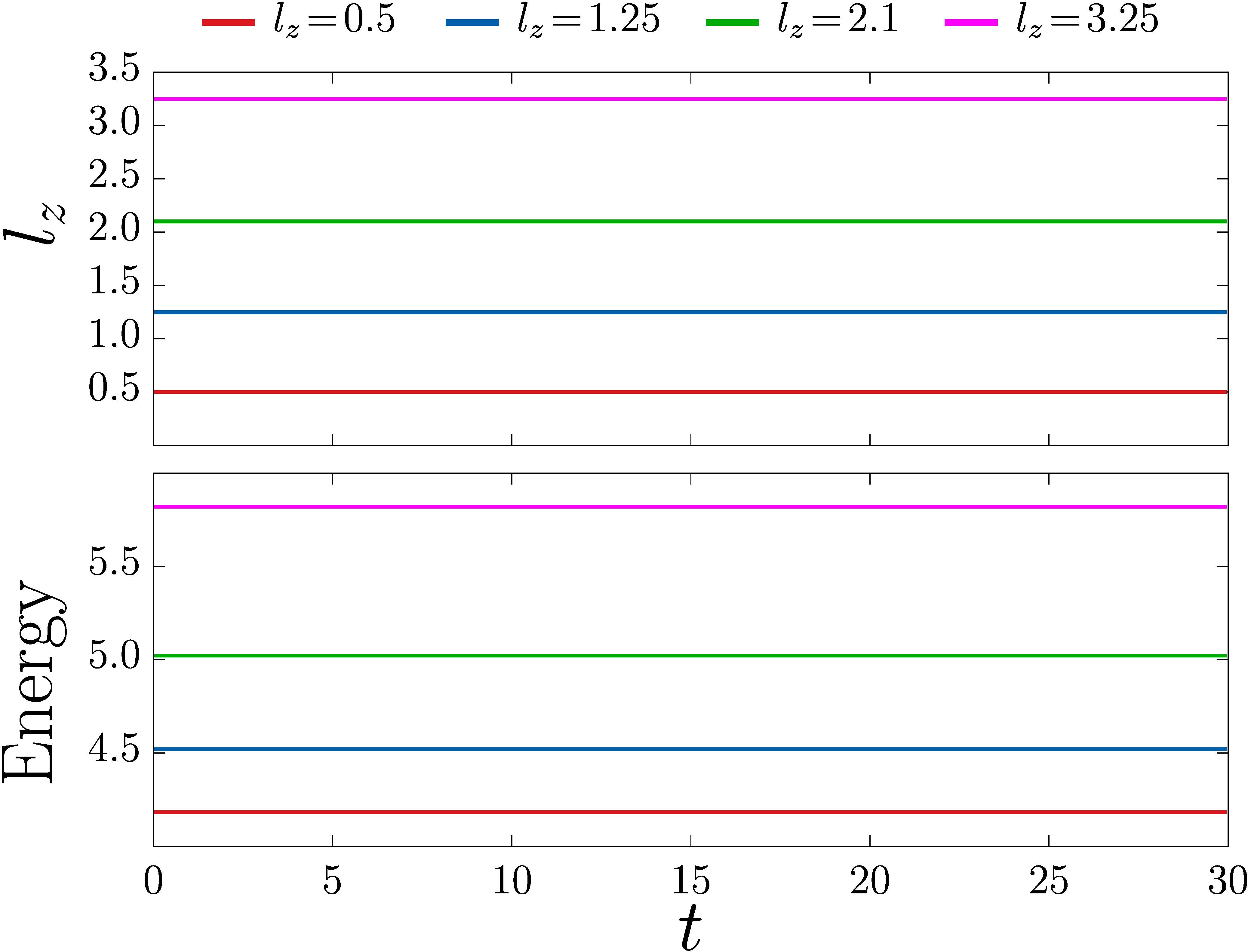}
\hss}
\caption{ 
Details of the results of simulations where the particle number is $N=1$. The dependence 
of the Lagrange multipliers on the constrained angular momentum $l_z$ are displayed on 
the leftmost panel. 
The right panels illustrate that the time-evolution algorithm \Eqref{Eq:TDGP:OP} 
used here indeed preserves the energy (bottom), the angular momentum (top) and 
the particle number (not shown).
}
\label{FigApp:Vortex:1}
\end{figure}

Numerical minimization of the weak formulation \Eqref{Eq:Minim:1} of the augmented 
Lagrangian \Eqref{Eq:ALM:Hamiltonian} gives the minimal energy states under the 
specified values of the constraints \Eqref{Eq:ALM:Constraints}. In all generality 
we consider unit particle number $N=1$ for various values of the angular momentum 
$l_z$. The corresponding values of the Lagrange multipliers are displayed in 
figure~\ref{FigApp:Vortex:1}.

\subsection{Time evolution: forward extrapolated Crank-Nicolson algorithm}

To address the question of the time-dependent Gross-Pitaevskii equation, the 
strategy is to use a Crank-Nicolson algorithm \cite{Crank-1996} to iterate 
the time series. More precisely, for efficient calculations, we write a semi-implicit 
scheme where the nonlinear part is linearized using a forward Richardson extrapolation. 
The time-dependent Gross-Pitaevskii reads as:
\Equation{Eq:TDGP:0}{
i\partial_t\psi=-\frac{1}{2}\nabla^2\psi +\left[V(\x)+g|\psi|^2\right]\psi \,.
}
The weak from, obtained by multiplying by test functions $\psi_w\in\mathcal{H}(\Omega)$ 
and integrating by parts reads as, in terms of the inner products \Eqref{Eq:Inner}, 
\Equation{Eq:TDGP:1}{
\ScalarProd{\psi_w}{i\partial_t\psi} =\frac{1}{2}\ScalarProd{\Grad\psi_w}{\Grad\psi}
+\ScalarProd{\psi_w}{\left[V(\x)+g|\psi|^2\right]\psi}\,.
}
The discretization of the time turns the continuous evolution into a recursive series 
over the uniform partition $\lbrace t\rbrace_{n=0}^{N}$ of the time variable. The time 
variable is discretized as $t=n\Delta t$ and the wave function at the step $n$ is 
$\psi_n:=\psi(n\Delta t)$. The Crank-Nicolson scheme, uses (Forward-)Euler definition 
of the time derivative, while the r.h.s of Eq.~\Eqref{Eq:TDGP:1} is evaluated at the 
averaged times. Introducing the notations 
\Equation{Eq:TDGP:Notations}{
\partial_t\psi :=\delta_t\psi_n = \frac{\psi_{n+1} - \psi_{n}}{\Delta t}
\,,~~\text{and}~~
\bar{\psi}_n = \frac{\psi_{n+1} + \psi_{n}}{2}\,,
}
the Crank-Nicolson scheme for the time-dependent Gross-Pitaevskii equation 
\Eqref{Eq:TDGP:1} reads as:
\Equation{Eq:TDGP:2}{
\ScalarProd{\psi_w}{i\delta_t\psi_n} =
\frac{1}{2}\ScalarProd{\Grad\psi_w}{\Grad\bar{\psi}_n}
+\ScalarProd{\psi_w}{V(\x)\bar{\psi}_n}
+\ScalarProd{\psi_w}{g|\bar{\psi}_n|^2\bar{\psi}_n}	\,.
}
This fully implicit scheme results in a nonlinear algebraic system which is very 
demanding to solve. The alternative to solving the nonlinear algebraic problem is 
to approximate the nonlinear part in terms of values at previous time steps. Very 
schematically, the idea of the modified algorithm is to approximate the fields in the 
nonlinear term by using an extrapolation of the previous time steps, that retains the 
same order of truncation error as the rest of time series. Thus, using the forward 
extrapolation the averaged wave function in the non-linear term becomes 
$\bar{\psi}_n\approx(3\psi_{n}-\psi_{n-1})/2$. 
Next, defining the time-discretized operators: 
\SubAlign{Eq:TDGP:Operators}{
\Op_1\psi&=\ScalarProd{\psi_w}{\frac{i\psi}{\Delta t}} 
-\ScalarProd{\Grad\psi_w}{\frac{1}{4}\Grad\psi}
-\ScalarProd{\psi_w}{\frac{1}{2}V(\x)\psi}
\,, \\
\Op_2\psi&=\ScalarProd{\psi_w}{\frac{i\psi}{\Delta t}} 
+\ScalarProd{\Grad\psi_w}{\frac{1}{4}\Grad\psi}
+\ScalarProd{\psi_w}{\frac{1}{2}V(\x)\psi}
\,, \\%
\U_n\psi&=\ScalarProd{\psi_w}{\frac{g}{8}|3\psi_{n}-\psi_{n-1}|^2\psi}
\,,
}
allows to rewrite Eq.~\Eqref{Eq:TDGP:2} in the compact form
\Equation{Eq:TDGP:3}{
\Op_1\psi_{n+1}=\Op_2\psi_{n} + \U_n(3\psi_{n}-\psi_{n-1})\,.
}
Hence the time-evolution is formally given by the recursion 
\Equation{Eq:TDGP:OP}{
\psi_{n+1}=\Op_1^{-1}\big[\Op_2\psi_{n} + \U_n(3\psi_{n}-\psi_{n-1})\big]\,.
}
The spatial discretization is achieved by replacing the wave function $\psi$ by its 
finite-element space representation $\psi^{(h)}\in V_h(\mathcal{T}_h,\mathcal{P}^{(2)})$ 
in the time-dependent Gross-Pitaevskii equation \Eqref{Eq:TDGP:OP}. The test functions 
$\psi_w$ now take values in the same discrete space as $\psi^{(h)}$. Denoting the matrix 
representation of the time-discretized evolution operators \Eqref{Eq:TDGP:Operators}, as: 
\Equation{Eq:TDGP:Matrices}{
\Op_1\mapsto{\bs M}_{\psi} \,,~~~
\Op_2\mapsto{\bs N}_{\psi} \,,~~~\text{and}~~~~
\U_n(3\psi_{n}-\psi_{n-1})\mapsto{\bs R}_{\psi} \,, 
}
the recursion Eq. \Eqref{Eq:TDGP:OP} reduce to a linear system that read as: 
\Equation{Eq:TDGL:modLCN:Matrix}{
\left[{\bs M}_{\psi}\right]\left[\psi_{n+1}^{(h)}\right] 
-\left[{\bs N}_{\psi}\right]\left[\psi_n^{(h)}\right]
=\left[{\bs R}_\psi\right]	\,.
}
The vector ${\bs R}_\psi$ which is a function of $\psi_n^{(h)}$ and $\psi_{n-1}^{(h)}$, 
has to be recalculated at each step. The matrices ${\bs M}_{\psi}$ and ${\bs N}_{\psi}$, 
on the other hand, are constant matrices to be allocated just once and are in principle 
easily preconditioned. Finally the recursion is thus given by
\Equation{Eq:TDGL:modLCN:Solve}{
\left[\psi_{n+1}^{(h)}\right]=
\left[{\bs M}_{\psi}\right]^{-1}\left(
\left[{\bs N}_{\psi}\right]\left[\psi_n^{(h)}\right]
+\left[{\bs R}_\psi\right]
\right)
\,,
}
which requires recalculating the vectors ${\bs R}_\psi$ at each step and then 
multiplying by the inverse matrices.

The numerical simulations of the time-evolution algorithm \Eqref{Eq:TDGL:modLCN:Solve} 
representing the discretized evolution scheme \Eqref{Eq:TDGP:OP} accurately reproduces 
the intrinsic physical properties of the time-dependent Gross-Pitaevskii equation. 
Namely it preserves the conserved quantities like the energy, the angular momentum, 
and the particle number. A consistency check, as displayed on figure \ref{FigApp:Vortex:1}, 
shows that these are indeed (exactly) conserved during the time-evolution.


\providecommand{\href}[2]{#2}\begingroup\raggedright\endgroup

\end{document}